\documentclass{aa}
\bibpunct{(}{)}{;}{a}{}{,}
\usepackage{latexsym}
\usepackage{natbib}
\usepackage{graphicx}
\usepackage{txfonts}
\usepackage{color}
\newcounter{IonCS}
\renewcommand{\ion}[2]{\setcounter{IonCS}{#2}#1\,{\scshape{\roman{IonCS}}}}

\newcommand{\sect}[1]{Sect.\,\ref{#1}}
\newcommand{\sects}[1]{Sects.\,\ref{#1}}
\newcommand{\fig}[1]{Fig.\,\ref{#1}}
\newcommand{\figs}[1]{Figs.\,\ref{#1}}

\newcommand{\equ}[1]{Eq.\,(\ref{#1})}
\newcommand{\ang}[1]{#1\,\AA}
\newcommand{\abs}[1]{\lvert #1\rvert}

\graphicspath{{./}{fig_eps/}{fig_pdf/}}

\begin{document}

%
\title{A model for the formation of the active region corona driven by magnetic flux emergence} 

\titlerunning{Formation of an active region corona}
\authorrunning{Chen et al.}

\author{F.~Chen\inst{1}, H.~Peter\inst{1}, S.~Bingert\inst{1}, M.~C.~M.~Cheung\inst{2}}

\institute{Max Plank Institut f\"ur Sonnensystemforschung, Justus-von-Liebig-Weg 3,
           37077 G\"ottingen, Germany\\
           email: chen@mps.mpg.de
           \and
           Lockheed Martin Solar and Astrophysics Laboratory, 3251 Hanover St, Palo Alto, CA 94304, USA
}

\date{Received ... / Accepted ...}

\abstract%
%
{}
{%
We present the first model that couples the formation of the corona of a solar 
active region to a model of the emergence of a sunspot pair. This allows us to study
when, where, and why active region loops form, and how they evolve.
}
{%
We use a 3D radiation MHD simulation of the emergence of an
active region through the upper convection zone and the photosphere as a
lower boundary for a 3D MHD coronal model. The latter accounts for
the braiding of the magnetic fieldlines, which induces currents in the corona
heating up the plasma. We synthesize the coronal emission for
a direct comparison to observations. Starting with a basically field-free
atmosphere we follow the filling of the corona with magnetic
field and plasma.
}
{%
Numerous individually identifiable hot coronal loops form, and reach temperatures 
well above 1 MK with densities comparable  to observations. The footpoints of these loops are found where small patches of magnetic flux concentrations move into the sunspots. The loop formation is triggered by an increase of upwards-directed Poynting flux at their footpoints in the photosphere. In the synthesized EUV emission these loops develop within a few minutes. The first EUV loop appears as a thin tube, then rises and expands significantly in the horizontal direction. Later, the spatially inhomogeneous heat input leads to a fragmented system of multiple loops or strands in a growing envelope. 
}
{}

\maketitle

\section{Introduction\label{S:intro}}

In the process of emergence of magnetic field through the surface of the Sun a group of sunspots can form and the upper atmosphere is heated and filled with plasma. This forms  coronal loops prominently visible in extreme ultraviolet (EUV) and X-ray observations. In the early phase of theoretical investigations of  coronal loops, emphasis was put on their properties in an equilibrium state, i.e. once they have fully developed. For example, this led to the widely used scaling laws of \cite{Rosner+al:1978} relating energy input and loop length to temperature and pressure of a static loop. While now time-dependent models including plasma flows are widely used, they mostly concentrate on the loop evolution in a magnetic field that changes only slightly or not at all \citep[see e.g.\ review by][]{Reale:2010}. This applies to 1D models for the dynamic evolution of loops, e.g.\ accounting for siphon flows \citep{Boris+Mariska:1982}, in response to intermittent nanoflare heating \citep{Hansteen:1993}, or concerning catastrophic cooling \citep{Muller+al:2003}. Also 3D models of the coronal loop structure considered either a constant or slowly changing magnetic field in the photosphere \citep{Gudiksen+Nordlund:2002, Bingert+Peter:2011,Lionello+al:2013}. In all these models, the loops form and evolve in a preexisting coronal magnetic field above a well developed active region. However, during the early formation stage of an active region, when the first coronal loops form,  flux emergence is still significant. Using a magneto-frictional approach, \cite{Cheung+DeRosa:2012} followed the whole evolution of an active region. In particular, they determined the coronal magnetic field  in response to the flux emergence and used proxies based on  currents to obtain a rough estimate of the coronal emission. The goal of our study is to go one step beyond and follow the formation of the first loops in an emerging active region in a self-consistent model, properly accounting for the plasma properties and the magnetic field. This requires the treatment of an energy equation including radiative losses, heat conduction, and plasma heating. The synthesized coronal emission can be directly compared to observations.

Over the
last decade 3D models showed that in the magnetically closed {\em active region} corona the heat input most probably is related to the tangling of the magnetic field in the photosphere, either through fieldline braiding by the horizontal motions of granulation \citep{Parker:1972, Parker:1983} or flux-tube tectonics \citep{Priest+al:2002}. In both cases currents are induced in the corona and are subsequently dissipated.  In a numerical experiment \citet{Galsgaard+Nordlund:1996} could show that the driving of the magnetic field from the photospheric boundaries can indeed lead to significant build-up of currents to heat the corona through Ohmic dissipation. In a more realistic setup \citet{Gudiksen+Nordlund:2002,Gudiksen+Nordlund:2005a,Gudiksen+Nordlund:2005b} showed that a million K hot loop-dominated corona forms above an active region. They used an observed magnetogram and the fieldline braiding was driven by artificial granulation-like
motions at their bottom boundary. The induced currents $j$ are then converted to heat through Ohmic dissipation $H_{\rm{Ohm}}\propto\eta{j^2}$, where $\eta$ has a functional dependence on $j$. Transition region EUV emission lines synthesized from these models match major observed features such as the transition region redshifts  \citep{Peter+al:2004,Peter+al:2006} or temporal variability \citep{Peter:2007}. Using a constant value for $\eta$, \citet{Bingert+Peter:2011} studied the temporal variability of the heat input in detail. Such models show that the energy is predominantly deposited in quantities of about $10^{17}$ J \citep{Bingert+Peter:2013}, consistent with the predictions of nanoflares by \cite{Parker:1988}. Most recently, observed coronal structures could be reproduced by the 3D coronal models based on the observed magnetic field and flows in the photosphere alone \citep{Bourdin+al:2013}, and the constant cross section of loops was reproduced \citep{Peter+Bingert:2012}. While the above models are all based on DC-type Ohmic heating, recently \citet{Ballegooijen+al:2011} studied the AC energy input by Alfv\'en wave turbulence in a coronal magnetic flux tube using a reduced MHD approach and found it sufficient to heat the hot coronal loops.

On smaller spatial scales, more representative for network or plage patches, \citet{Martinez+al:2008, Martinez+al:2009}
 studied the process of a twisted flux tube emerging from the convective layer into
the lower corona.
In these realistic models the radiative transfer in the lower atmosphere is treated as well as the heat conduction and radiation needed in the coronal part. Descriptions of the flux tube emergence into the corona on larger scales, more representative of an active region, had to omit the realistic description allowing for the synthesis of photospheric and coronal emission \citep[e.g.][]{Fan:2001, Abbett+Fisher:2003, Manchester+al:2004, Archontis:2004,Magara:2006}. \citet{Hurlburt+al:2002} used a 2D axisymmetric magneto-convection sunspot model to construct the coronal magnetic field above. In a collection of 1D static models along the coronal fieldlines they then derived the EUV and soft X-ray emission and compared it to observations. This model is still quite idealized in the sense that the corona is described by a series of 1D equilibrium models along  magnetic fieldlines, and the heat input is determined by a (freely chosen) fraction of the Poynting flux that is generated by the magneto-convection.

In recent years it became possible to study the fine structure of sunspots \citep{Heinemann+al:2007} and simulate a whole sunspot and its surrounding granulation in 3D models from the convection zone to the photosphere \citep{Rempel+al:2009.sci}. This paved the road to investigate the emergence of a flux tube emerging through the surface resulting in a sunspot pair that is at the heart of an active region \citep{Cheung+al:2010}. These models are realistic in the sense that they include all the relevant physics, in particular the radiative transfer, in order to produce synthetic images from the simulation that closely resemble actual observations. A variant of this latter model (see \sect{S:model_flux}) will provide the lower boundary for our model of the corona above an active region.     

The goal of our study is to use a realistic model of the emergence of a sunspot pair in the photosphere as a lower boundary to drive a coronal model of the active region corona above. A single model encompassing a whole active region from the interior to the upper atmosphere is not yet possible. Solving both the radiative transfer in the lower atmosphere and  the anisotropic heat conduction in the corona in a computational domain covering a full active region is too demanding in terms of computing resources. Instead, we couple our coronal model to the flux-emergence simulation by driving the coronal model at the bottom boundary with the results from the flux-emergence simulation in the (middle) photosphere.

 The model setup is described in \sect{S:model}, before we present the appearance of the model corona in \sect{S:appear}. A detailed analysis along the coronal loop that forms first during the flux emergence is given in \sect{S:loop} followed by a discussion of the  three-dimensional nature of the loop in \sect{S:sect}. Finally we discuss the trigger of the coronal loop formation in \sect{S:trigger}.

\section{Model setup\label{S:model}}
\subsection{Flux emergence simulation\label{S:model_flux}}

The flux-emergence simulation we use to determine the lower boundary of our coronal model is based on the numerical experiment by \cite{Cheung+al:2010}. Here we actually use a variant with a slightly larger computational domain and without imposed twist for the emerging flux tube \citep{Rempel+Cheung:2013}.  In that model, the fully compressible MHD equations are solved including  radiative transfer and a
realistic (tabulated) equation of state. The computational domain covers 147.5$\times$ 73.7 $\times$16.4\,Mm$^3$, where the optical depth of unity is found about 15.7\,Mm above the bottom. A torus-shaped magnetic flux tube containing 1.7$\times$10$^{22}$\,Mx is kinematically advected into the domain through the bottom. The flux tube is then buoyantly rising through the upper convection zone and breaks
into photosphere.

In the early stage of the flux emergence simulation, the flux tube is still
below the photosphere and the photosphere is basically field-free.
While the flux tube is further rising, first a lot of small-scale flux concentrations emerge on the photosphere. Later, larger flux concentrations form through the coalescence of small magnetic flux patches.  In this process the granular motions play an important role. In the later stage of the simulation, some 27\,h after the start of the simulation, a pair of sunspots forms in the photosphere. The diameter of  each spot is about 15\,Mm. The magnetic field strength in the center of the spots exceeds 3000\,G, which is compatible with that of a large active region on the Sun.
The total duration of the simulation was about 30\,h.

We extract the quantities in the photosphere ($\tau{=}1$) of the simulation and use them to drive the coronal model (\sect{S:model_coupling}).

\subsection{Coronal model\label{S:model_corona}}
To model the corona above the emerging active region we employ a 3D magnetohydrodynamics (MHD) simulation.  Vertically this stretches from the photosphere into to the corona. Our model follows the investigations by \citet{Bingert+Peter:2011, Bingert+Peter:2013}. We solve the induction equation together with the  mass, momentum, and energy balance. We account for heat conduction along
the magnetic fieldlines \citep{Spitzer:1962}, radiative losses in the corona
\citep{Cook+al:1989}, and Ohmic heating, so that the coronal pressure is set self-consistently. This allows us to synthesize the coronal emission to be expected from the model domain. As proposed by \citet{Parker:1983}, magnetic fieldline braiding by photospheric motions induces currents in the corona, which are converted to heat by Ohmic dissipation.
As in \citet{Bingert+Peter:2011, Bingert+Peter:2013}, we use a constant resistivity $\eta$. We set $\eta$ by requiring the magnetic Reynolds number to be of order unity using the grid spacing as the length scale. This implies that $\eta$ is $10^{10}$\,m$^2$/s. To numerically solve the MHD equations we use the Pencil code
\citep{Brandenburg+Dobler:2002}.\footnote{See also http://pencil-code.googlecode.com/.} Further details on the  model can be found in \cite{Bingert+Peter:2011}.

The domain of the coronal model matches that of the flux emergence simulation in the horizontal direction, i.e.\ it spans 147.5$\times$73.7\,Mm$^2$. This is resolved by 256$\times$128 grid points, implying a 576\,km equidistant grid spacing. In the vertical direction the domain spans 73\,Mm above the photosphere. In the vertical direction we implement a stretched grid with 256 points. The vertical grid spacing near the bottom is about 32\,km, (matching the flux-emergence model), and in the coronal part it is about 300\,km.

The initial atmosphere is determined by a temperature stratification similar to that of the Sun, which increases from 5100\,K in the photosphere to 1\,MK in the corona with a smooth transition at about 3\,Mm above the photosphere. The initial density is calculated from hydrostatic equilibrium, with the  density at the bottom boundary matching the average density of the extracted layers from the flux emergence simulation. The initial magnetic field is  a potential field extrapolated from the photospheric magnetogram derived from  the flux emergence simulation (see \sect{S:model_coupling}).

The lateral boundaries are periodic and the top boundary is closed, with a zero temperature gradient. The bottom boundary
is set to drive the coronal model by the flux emergence simulation
(\sect{S:model_coupling}). Our previous models \citep[e.g.][]{Bingert+Peter:2011,Bingert+Peter:2013,Bourdin+al:2013}
used an observed magnetic field and an observed or generated velocity field to prescribe the bottom boundary conditions, which was then driving the model. Here we  impose the properties from the flux-emergence simulation in a similar fashion.

\subsection{Coupling of the flux emergence and coronal model\label{S:model_coupling}}

To drive the coronal model by the separate flux-emergence simulation, we
have to couple the two simulations.
This is done by specifying the bottom boundary condition of the coronal model, viz.\ the ghost layers and the bottom layer of the physical domain. For this we extract the density, temperature, and magnetic field in four consecutive layers from the photosphere in the flux-emergence simulation at a height of the average level of unity optical depth. We use a nearest-neighbour method to spatially interpolate the values in the horizontal direction to match the grid of the coronal simulation. Unfortunately, a lot of the kinetic energy in the small scales (${\approx}70\%$ of the total kinetic energy) is lost during the mapping. The impact of this on the coronal structure is discussed later in \sect{S:trigger}. The vertical spacing of the coronal model at the bottom and the flux-emergence simulation is identical. We store these values for the boundary and ghost layers with a cadence of 25\,s which is more than sufficient to follow the emergence of the small flux elements in-between the granulation. At each time step of the coronal model when updating the boundary conditions, we feed the extracted values into the boundary and ghost layers of the coronal model. In this process we apply a linear interpolation in time between the boundary data stored with 25\,s cadence.

Because no significant amount of magnetic flux is found in the photosphere before about 21\,h, and the spots do not start to form before 24\,h into the flux-emergence simulation, we start the coronal model at 21\,h after the start of the flux-emergence simulation. This should give enough time before magnetic concentrations form which are strong enough to give rise to the formation of coronal loops. The driving at the bottom boundary is switched on gradually within the first few minutes after the start. We also apply a damping to the velocity in the first 15\,min to dissipate shocks generated from the mismatch between the initial condition and the bottom boundary. The coronal model is evolved for half a solar hour, with a basic energy equation without energy input, loss, or conductive transport, in order to relax from the initial condition. Then anisotropic heat conduction, radiative losses through optically thin radiation, and Ohmic and viscous heating are switched on. The coronal model evolves for about 8 solar hours almost to the end of the flux-emergence simulation, until the simulated spots are fully formed. 

The main goal of this study is to investigate the initial formation of a coronal loop in an emerging active region. Therefore,  in this paper we  concentrate  on a time span of about 30\,min after the first coronal loop becomes clearly visible in the synthesized EUV emission, which is about 25.5\,h after the start of the flux-emergence simulation.

\begin{figure*}
\includegraphics[width = 18 cm]{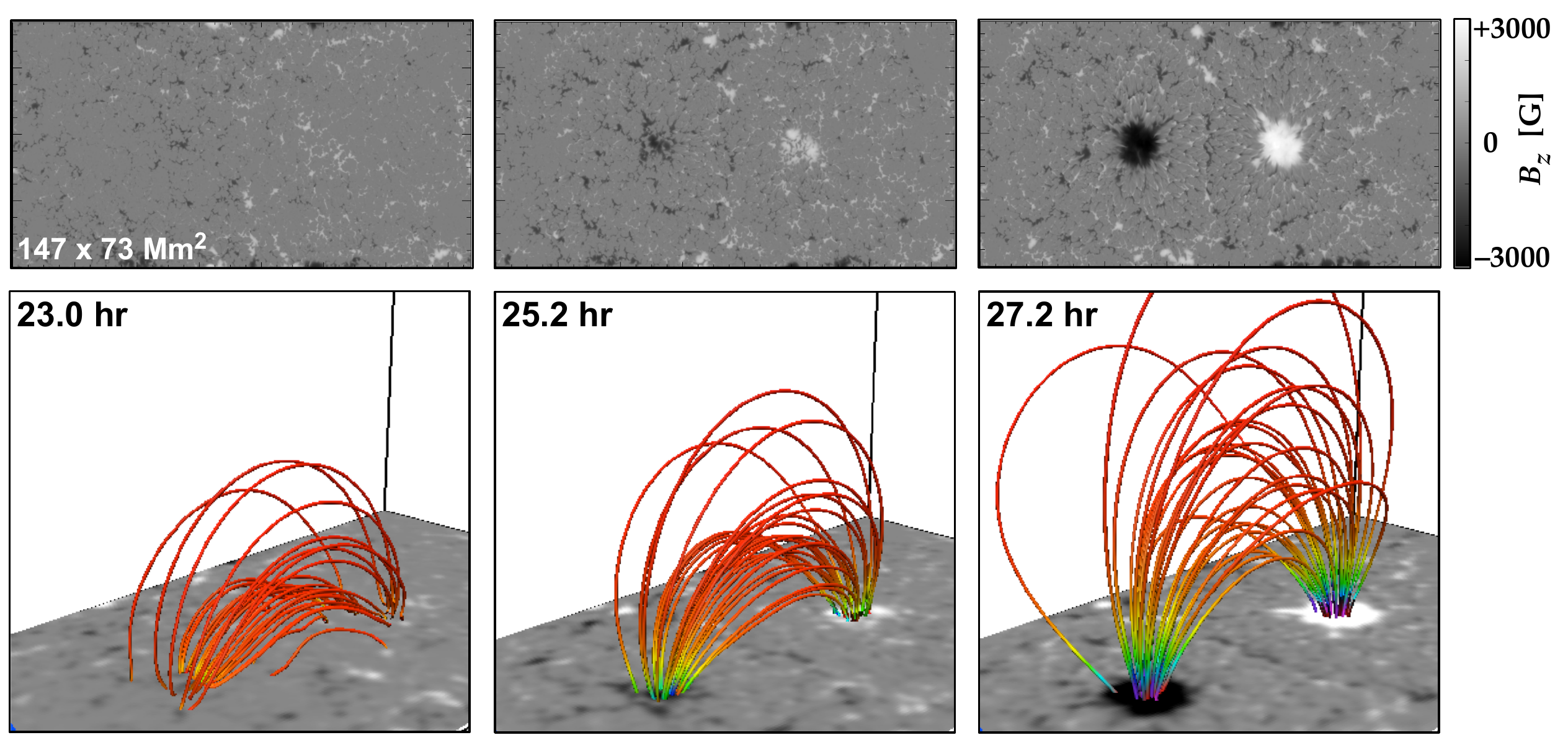}
\caption{Overview of evolution of magnetic field. The top row shows  the magnetogram at the bottom boundary in the photosphere at three different times (vertical magnetic field). The horizontal
extent in the top panels is 147.5$\times$73.7\,Mm$^2$. The bottom row shows the evolution of a group of magnetic fieldlines in the coronal model driven by the flux emergence. The color coding on the fieldlines shows the magnetic field strength (red is low and purple is high). The bottom of the 3D rendering boxes
are the same magnetogram as the top row. The times in the panels refer to the time since the start of the flux-emergence simulation. See \sect{S:appear_mag}.   
\label{F:3d}}
\end{figure*}

\section{Coronal loops appearing in an emerging active region\label{S:appear}}
\subsection{Magnetic expansion into the corona\label{S:appear_mag}}

Driven by the magnetic flux emergence through the bottom boundary, the magnetic field expands into the corona. In \fig{F:3d} we show an overview of the evolution of the magnetic field in the coronal simulation over four hours. We select the fieldlines at an early stage of the simulation by random seeds in a small volume in the lower middle of the computational box and trace their evolution.%
\footnote{The algorithm for fieldline tracing used by VAPOR with the results shown in \fig{F:3d} is described at http://www.vapor.ucar.edu/docs/vapor-renderer-guide/flow-tab-field-line-advection. It assumes that fieldlines are frozen in the plasma elements (i.e., infinite conductivity) everywhere  and follows the motions of plasma elements. The influence of magnetic diffusivity is discussed in \sect{S:loop}.} Thus the lines in the three lower panels of  \fig{F:3d} show the same set of fieldlines and how they evolve in time. 

At 23.0\,h, there are already lots of small-scale flux-concentrations in the photosphere. Low-lying fieldlines connect these small
elements (not shown in the figure). As discussed in \cite{Cheung+al:2010}, these small flux concentrations are part of the flux tube brought to the surface through the near-surface convection. Because of
the large scale of the emerging flux tube, the large-scale magnetic connections in the figure (at 23\,h) show a bipolar pattern. 

After two hours evolution at about 25.2\,h,  more magnetic flux emerged through the photosphere and the small-scale flux-concentrations begin to coalesce. Now the large-scale magnetic field concentrations start to become visible in the photosphere. This is also illustrated by the fieldlines whose footpoints are moving closer to each other now concentrating near the simulated spots. The magnetic field strength near the footpoints increases. At the same time, the fieldlines expand upward into the higher atmosphere.

After another two hours, around 27.2\,h, a pair of simulated spots, where the magnetic field strength is over 3000 G, has formed in the photosphere. Now at the end of the coalescence process the footpoints of the fieldlines are bundled closely together. The central part of the set of fieldlines continues to expand into the higher corona.

The  evolution of the magnetic field at the bottom boundary of our coronal model follows that in the flux emergence simulation, of course, albeit at a reduced spatial resolution. While most of the fine structures are lost due to the lower resolution, the photospheric magnetic field still captures the formation of sunspots by coalescence of small (down to the resolvable scale) flux elements.

\subsection{Appearance of a coronal loop\label{S:appear_loop}}
One of the key interests of this study is whether coronal loops will form during the active region formation. Here and in most of the cases in this paper  the term {\em coronal loop} refers to a loop-like structure identifiable in (real or model-synthesized) EUV observations of the corona. Whenever we refer to the magnetic field that confines the plasma contributing to the EUV loop emission, we always use the term {\em magnetic tube}.

To perform a direct comparison between our model and observations, we synthesize EUV images using the AIA response function \citep{Boerner+al:2012}, following the procedure of \citet{Peter+Bingert:2012}. Here we concentrate at the AIA \ang{193} channel. It looks similar but not identical to \ang{171} and \ang{211} channels, which sample the 1 to 2\,MK plasma, too. The \ang{193} and \ang{211} channels have also contributions from cooler plasma, in particular in quiet regions. However, this does not play a major role in our active region (model).

To form a coronal loop visible in EUV, one has to bring up enough plasma into the upper atmosphere along a fieldline  and heat it to at least $10^{6}$\,K. In our model, the heating is by dissipation of currents  which are induced by braiding of magnetic fieldlines through photospheric motions. If the magnetic field strength in the photosphere is low, the plasma motion can braid the magnetic fieldlines efficiently, but the induced currents will be weak, too. On the other hand, the braiding does not work in very strong magnetic flux concentrations, because there the plasma motions are suppressed. Both moderately high magnetic field and horizontal velocity are needed to induce enough current. This favorable combination is found at the periphery of sunspots, and in  particular in our model after some  25.5\,h after the start of the flux-emergence simulation.
Therefore all times (usually given in minutes) mentioned in the remainder of the paper will refer to this time, i.e., in the following $t{=}0$ refers to 25.5\,h after the start of the flux emergence simulation.  

\begin{figure*}
\sidecaption
\includegraphics{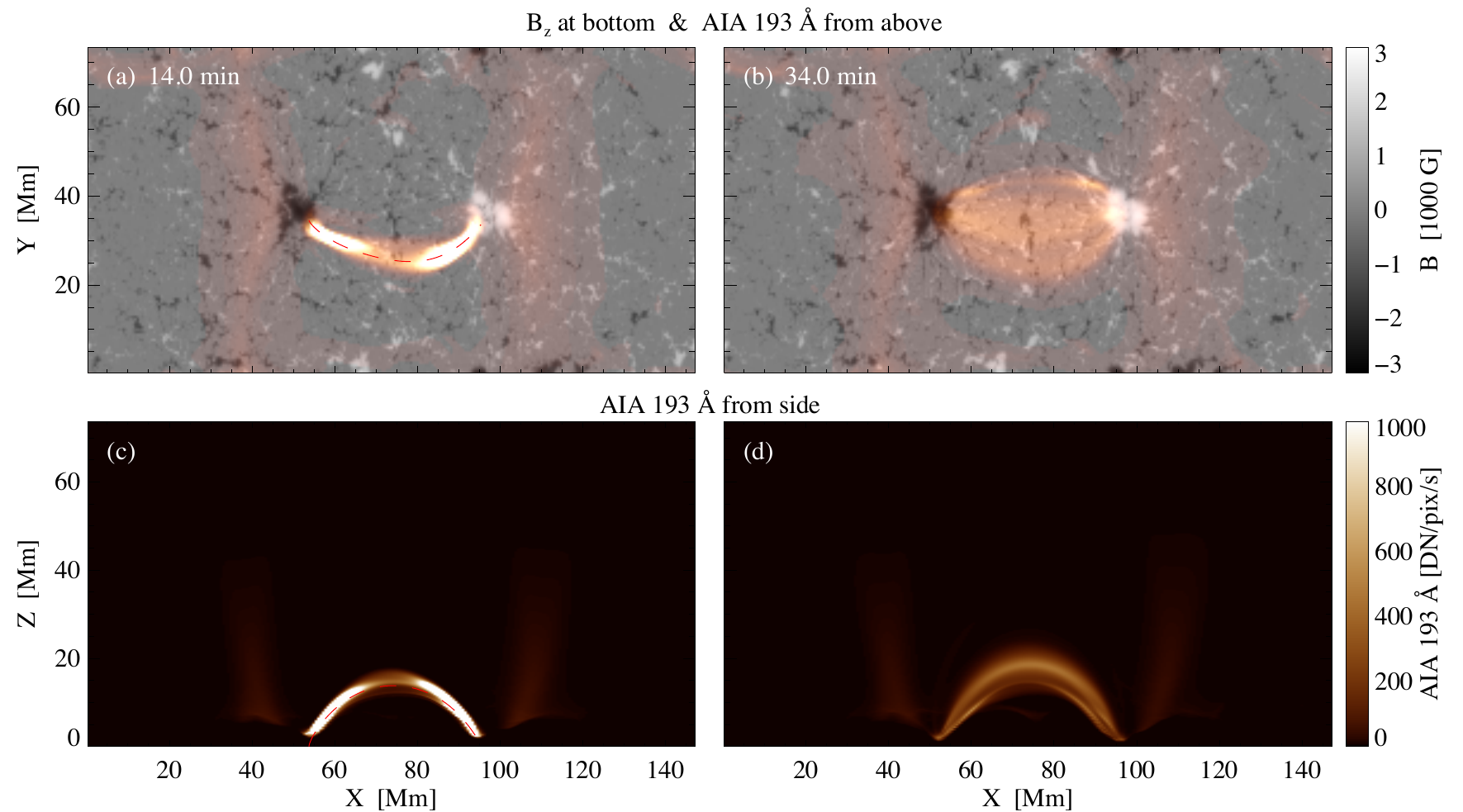}
\caption{Evolution of synthesized coronal emission and magnetic field. The top panels show photospheric magnetograms (vertical component), overlaid with the synthesized coronal emission as it would be seen in the  AIA \ang{193} channel. The bottom panels  show synthesized AIA \ang{193} images as seen from the side. The synthetic emission is integrated along the line of sight , comparable to what is seen at disk center (top) or the limb (bottom). The two columns show snapshots  from the simulation 20 minutes apart. Times refer to 25.5\,h after the start of the flux-emergence simulation. The dashed line in the left panels shows the fieldline at the spine of the loop selected for the analysis in \sect{S:loop}. An animation of the figure is available in the online edition and at \newline\protect\url{http://www2.mps.mpg.de/projects/coronal-dynamics/res/emerging_ar_f1.mp4}. See \sect{S:appear_loop}.
\label{F:em}}
\end{figure*}

Once the energy input into the corona is sufficient, EUV loops will start to form.  In \fig{F:em} we show the synthesized images for the AIA \ang{193} channel integrated along the vertical and the horizontal coordinate at two different times 20 minutes apart. These views correspond to observations near the disk center (top panels) and at the limb (bottom panels). The left column shows a snapshot at $t{=}14$\,min, just after the first EUV loop appears.
At this time we see a single EUV loop forming, at later times more loops form at different places. Here we concentrate on the first single EUV loop in order to better isolate the processes triggering its formation.\footnote{In a more recent not yet fully finished numerical experiment we see also multiple loops forming at this early stage, so the limited spatial resolution in this model plays a role, too.}
The right panels of \fig{F:em} show the coronal emission at $t{=}34$\,min, after the loop started to fragment into several individual loops (see \sect{S:sect_frag}). From almost no emission to clearly detectable count rates in the synthesized images it
takes only ${\approx}5$\,min (see  the animation with \fig{F:em}). In this paper we will mainly concentrate on the initial evolution of the loop system during about 15\,min. 

The EUV loop is rooted in the periphery of the simulated spots, which is clear from the top panels of \fig{F:em} showing an overlay of magnetogram and emission. This is consistent with the long-known observation fact that the footpoints of coronal loops are not in the umbra at the higher field strengths, but in the periphery, the penumbra \citep{Bray+al:1991}. Even though the flux emergence simulation does not contain a proper penumbra \citep{Cheung+al:2010} it is clear that the loops are rooted in a region where convection can do considerable work to the magnetic field in the photosphere (see \sect{S:trigger}) and thus induces strong currents in the corona.

The visible top of the EUV loop rises upwards by 10\,Mm within 20\,min, which means a 10\,km\,s$^{-1}$ average upward velocity of the apex. The cross section of the EUV loop expands in the vertical direction during this rise. However, the EUV loop expands even more strongly in the horizontal direction after its initial appearance as a relative thin tube of up to 5\,Mm diameter and  45\,Mm length. Finally the emission covers the whole area in-between the two simulated spots, with a few fragments in the relatively diffuse loop emission (\fig{F:em}b). 

The above discussion, in particular the late fragmentation, shows that  a 3D treatment of the loop formation is essential. Still, in the early phase, the loop evolution appears to be close to a single monolithic loop. Therefore, we first analyze the 1D evolution along the spine of the emerging loop in \sect{S:loop}. The full 3D aspects and the trigger of the loop formation will be addressed after that in \sects{S:sect} and \ref{S:trigger}.
 
\section{The 3D loop collapsed to one dimension\label{S:loop}}
In the solar corona, the high electric conductivity  prevents slippage of the fully ionized plasma across the field. The dominant magnetic energy assures that the Lorentz force determines motions perpendicular to the magnetic field. The anisotropic heat conduction quickly spreads the thermal energy along the fieldlines. Under such circumstances, often a simplified 1D model along a magnetic fieldline is sufficient to describe a coronal loop, although EUV loops are 3D structures in nature, as we show in this study.

In a 3D model, one can analyze the dynamics and thermal structures along a certain magnetic fieldline, which should then give results equivalent to that of a 1D loop model. For the loop forming in our 3D model, this applies in the early phase, when the loop is still confined to a thin magnetic tube. This assumption breaks down in the later phase, when the loop fragments into several substructures (see \sect{S:sect}). 

The spine of a EUV loop, which can be considered as being the central magnetic fieldline in the structure, is assumed to be static in most 1D models. However, it evolves in a self-consistent manner in 3D models. For 3D models of mature active regions \citep{Gudiksen+Nordlund:2005a, Gudiksen+Nordlund:2005b, Bingert+Peter:2011}, the magnetic field evolution follows the shuffling of  footpoints of the fieldlines by granular motions and the change in morphology is very gentle. It is  quite different in our model. When the first coronal loop becomes visible, the flux emergence is still going on, and the coronal magnetic field changes dramatically. To analyze the evolution of an equivalent 1D model, we need to follow the magnetic fieldline in time and extract all physical quantities along this evolving fieldline. 

During this tracing, we assume that magnetic fieldlines are frozen-in with the plasma elements, as it should be in the case of high electric conductivity. Although there is a constant numerical resistivity in the induction equation in our simulation, the typical diffusion speed over 10\,Mm is of the order of 1\,km s$^{-1}$, which is smaller than the typical velocities (perpendicular to $B$) associated with the expansion of the magnetic fieldlines. In practice, we first follow the motion of the plasma element at the apex of a magnetic fieldline, and then calculate the new fieldline passing through the new position of this plasma element.

\subsection{Thermal structure and dynamics of the loop\label{S:loop_dyn}}

\begin{figure*}
\includegraphics{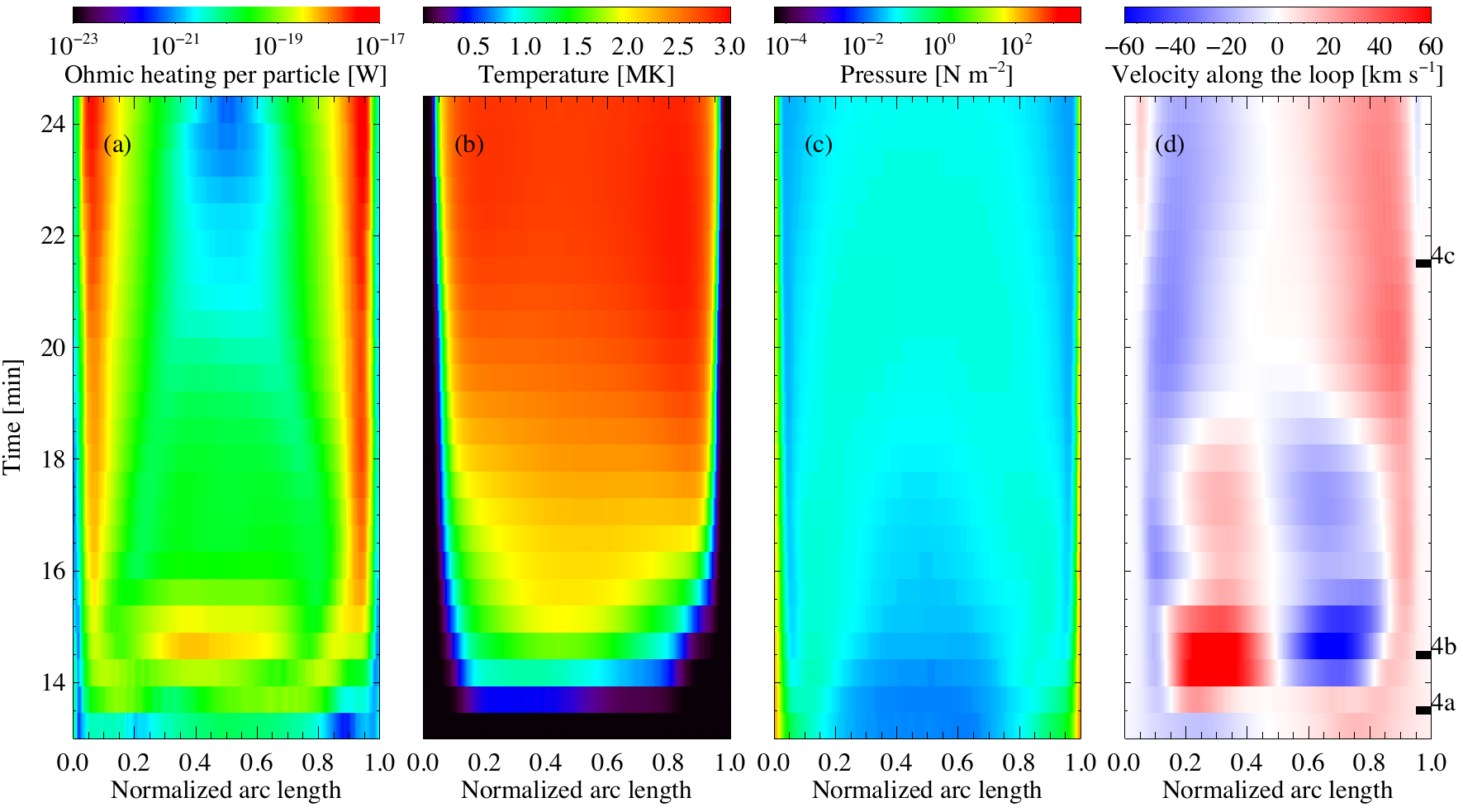}
\caption{Space-time diagram of  emerging loop. All properties are shown as functions of normalized arc length along the fieldline at the spine of the forming loop and time. Because the fieldline is followed in time, its length is changing and the arc length along the fieldline is normalized to the length at the respective time.  The loop footpoints are at arclengths 0 and 1. In the velocity plot positive velocities (red) indicate a flow in the direction of arc lengths from 0 to 1 (``to the right''). In the color scale for the temperature green roughly corresponds to the peak contribution to the AIA \ang{193} channel.  The marks 4a, 4b, and 4c in the right panel indicate the times shown in the  three panels of  \fig{F:ener}. See \sect{S:loop_dyn}.
\label{F:loop}}
\end{figure*}


We choose a fieldline along the spine of the loop seen in the AIA \ang{193} image at $t{=}14$\,min, when the loop is clearly defined (left panels of \fig{F:em}). In \fig{F:loop} we show a space-time diagram for this fieldline traced following the strategy above. The coordinate along the magnetic fieldline is normalized by the instantaneous length integrated between the two photospheric footpoints of the magnetic fieldline.  During the time we investigate the evolution of the loop (approx.\ 14\,min to 24\,min) the fieldline at its spine lengthens by some 10\% to 15\%.

In the very early stage, the plasma along the fieldline is still cold and the pressure near the top is low. In this early phase ($t{<}14$\,min), there is some weak draining along the fieldline, due to the slow rise of the fieldline driven by the flux emergence.

At $t{\approx}14$\,min, Ohmic heating increases through the whole loop, and the coronal temperature quickly increases to over 1 MK (\fig{F:loop}a, b). Here we analyze the temporal change of the heating, and discuss in \sect{S:trigger} the self-consistent trigger of the increase of the heating rate in the 3D model. Very efficient heat conduction along the loop ensures an almost constant temperature along the fieldline in the coronal part (\fig{F:loop}b). At the same time, the heat conduction transfers the energy deposited in the corona down to the cold dense chromosphere and induces an evaporation upflow (\fig{F:loop}d). This flow fills the loop with plasma as is reflected by the increase of loop pressure in the \fig{F:loop}c.

On the particular fieldline we analyse here, the plasma starts to increase
its temperature at around $t{=}14$\,min. On other fieldlines (reaching slightly
larger apex heights) the heating sets in earlier. Thus some emission in the
       AIA \ang{193} channel can be seen already before $t{=}14$\,min.

In a later stage ($t{>}16$\,min), the Ohmic heating drops and the filling of the loop gradually ceases (\fig{F:loop}a, d). The pressure gradient at this moment is not sufficient to balance gravity and thus to maintain an equilibrium. As a result, the plasma starts to drain, as demonstrated by the downflows in \fig{F:loop}d after $t{\approx}20$\,min. The loop temperature, which is over 2.5\,MK after $t{\approx}19$\,min, is maintained for a long time. This is consistent with the long cooling times for these high temperatures \citep{Klimchuk:2006}, which is about 30 min. The energy evolution of the loop is further analyzed in detail in \sect{S:loop_ener}.

There is a local pressure peak in the lower part on each side of the loop from $t{=}$13.5\,min to 18\,min. These peaks result in both upward and downward pressure gradient forces, which drive the flows to the loop top and the loop feet. The temperature of the downflow is $10^4$\,K to $10^5$\,K, which corresponds to transition region temperatures; that of the upflow is $10^6$\,K, which corresponds to coronal temperature. This would cause the transition region lines (formed below 0.5\,MK) to be redshifted and the coronal lines to be blueshifted. Thus, this is consistent with observations \citep{Peter:1999full,Peter+Judge:1999} and in line with processes found by
\cite{Spadaro+al:2006}
in 1D models and by \citet{Hansteen+al:2010} in 3D models.

\subsection{Energetics in the emerging loop\label{S:loop_ener}}
To investigate the energy budget controlling the thermal structure of the loop and its dynamics, we analyze the change of the thermal energy. We do this for the same magnetic fieldline chosen in \sect{S:loop_dyn}. The conservation of thermal energy (see derivation in Appendix \ref{appen}) can be written as %
\def\bbb{\rule[-2.ex]{0pt}{1ex}}
\def\bba{\rule[-1.3ex]{0pt}{1ex}}
\begin{eqnarray}
\nonumber
\left(\frac{\partial e_{\rm th}}{\partial t}\right)_{s} & = &
\underbrace{\bbb-u_{\parallel} \left(\vec{b}\cdot \nabla \right) e_{\rm th}}_{\text (1)}  ~~ \underbrace{\bbb-~~\frac{\gamma}{\gamma-1} p \left(\vec{b} \cdot \nabla \right)u_{\parallel} }_{\text (2)}  
\\[1.3ex]
&& ~~ + \underbrace{\bba~~Q~~}_{\text (3a, 3b)}  \underbrace{\bba-~~L~~}_{\text (4)} ~~ \underbrace{\bba-~~\vec{b}\cdot \nabla q_{\parallel}}_{\text (5)} ~.  
\label{E:ener}
\end{eqnarray}
Here $e_{\rm th}$ is the thermal energy per unit volume, following $e_{\rm th} = p/(\gamma -1)$. $p$ is the plasma pressure, and $\vec{u}$ the velocity, $L$ denotes the radiative losses through optically thin radiation, $\vec{b}$ is the unit vector of the magnetic field, $u_{\parallel}=\vec{u} \cdot \vec{b}$ the velocity along the magnetic field,  and $q_{\parallel} = -\kappa_0 T^{5/2} (\vec{b} \cdot \nabla T)$ the heat flux along the magnetic field. Energy is added through $Q = Q_{\rm Ohm}+Q_{\rm visc}$, with the Ohmic heating (3a) $Q_{\rm Ohm} = \eta\mu_0 j^2$ and viscous heating (3b) $Q_{\rm visc}=2\rho\nu \vec{S}^2$, where $\vec{j}$ is the current and $\vec{S}$ is the rate-of-strain tensor.  In an equilibrium model, the time derivative and velocity would vanish, and the energy would be balanced between  heating (3), radiative losses (4) and heat conduction (5). In our time-dependent 3D model,  the loop never reaches an equilibrium. The advection of thermal energy along the loop is given by term (1) and the compressional work (combined with the change of $e_{\rm th}$ due to the compressibility of the plasma) by term (2) in the above equation. 

\begin{figure}
\includegraphics{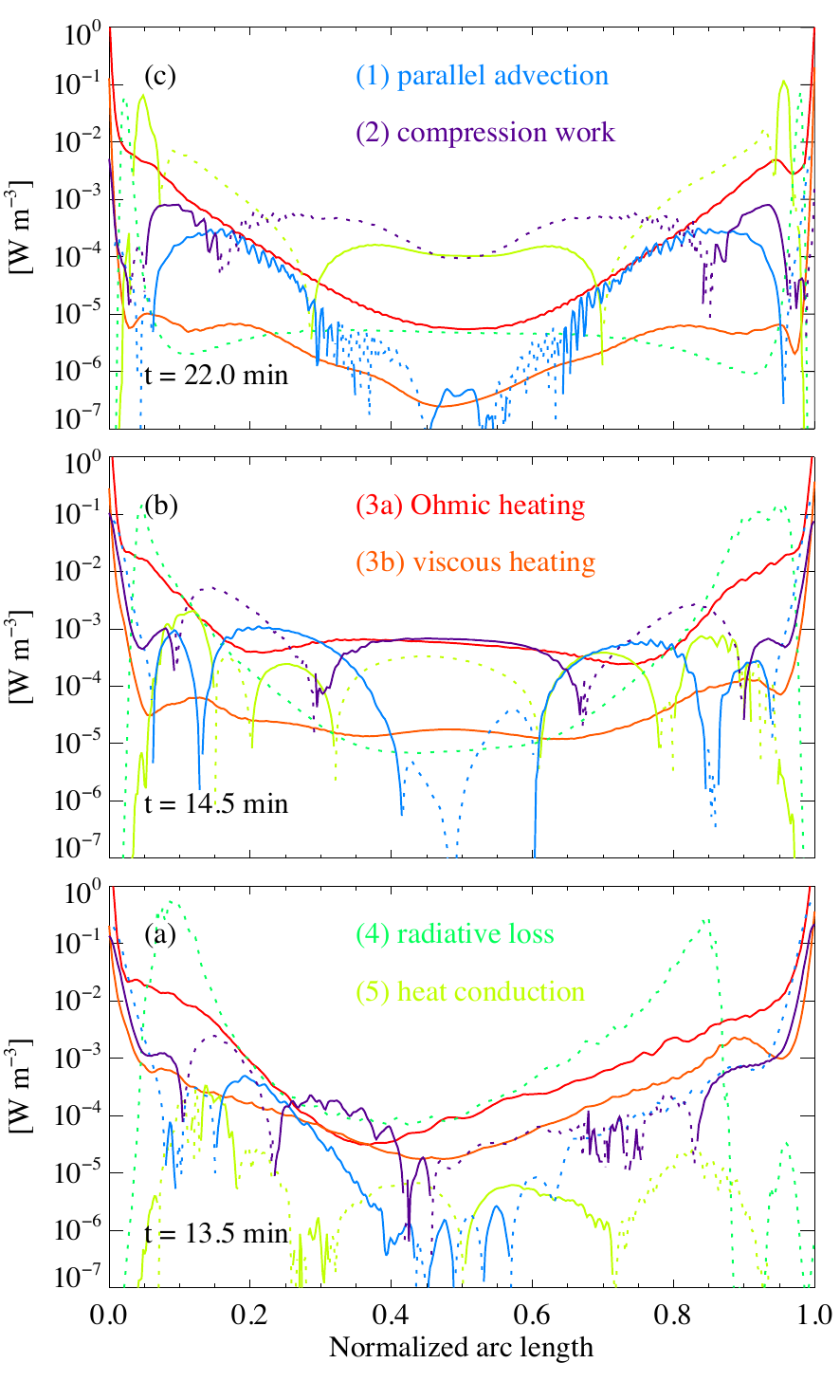}
\caption{Energy budget along the loop at three different times. The panels show snapshots during the phases of initiation (a), formation (b),
and cooling (c) at the time given with each panel. These times are indicated to the right of \fig{F:loop}d by the marks 4a, 4b, and 4c. The terms in   \equ{E:ener} are shown accordingly to the labels in the plots, the numbers correspondingly to those in   \equ{E:ener}. The line colors follow the same definition in all panels. Dashed lines indicate negative, solid lines positive  values. The arc  lengths are normalized, with 0 and 1 corresponding to the two footpoints of the fieldline in the photosphere. See \sect{S:loop_ener}. 
\label{F:ener}}
\end{figure}

In the following, we discuss the energy budget along the same fieldlines as in \sect{S:loop_dyn} before, during, and after the loop formation. For this we concentrate on the times $t{=}13.5$\,min, 14.5\,min, and 22.0\,min which are indicated in \fig{F:loop} by the marks 4a, 4b, and 4c, which refer to the respective panels in \fig{F:ener} showing the terms (1) to (5) in the energy budget along the fieldline.

\subsubsection{Initiation phase\label{S:loop_ener_i}}
In this early stage, there is a weak siphon flow in the loop (\fig{F:loop}d), which is probably driven by the stronger heat input near the left footpoint. Because the  loop is cool ($T{<}5 \times 10^4$\,K), the heat input is more or less balanced by radiative losses (\fig{F:ener}a). At this time the loop would be invisible in EUV images with count rates below the sensitivity (of AIA\ observations).  But at this moment, the Ohmic heating starts to increase. Although at the normalized arc length of 0.3 viscous heating is of the same magnitude as Ohmic heating, it is in general smaller than  Ohmic heating by at least one order of magnitude. Thus the increase of Ohmic heating is the primary cause of loop formation.

\subsubsection{Formation phase\label{S:loop_ener_f}}
At $t{=}14.5$\,min,  the Ohmic heating is high in the middle part of the fieldline, giving rise to the loop formation. Within less than a minute, the Ohmic heating rate has risen to a roughly constant level of almost $10^{-3}$\,W\,m$^{-3}$ in the hot coronal part (\fig{F:ener}b). Considering that the coronal part covers 20\,Mm to 30\,Mm along the loop, this implies an energy flux of ${\approx}10^4$\,W\,m$^{-2}$ into the loop, which would be consistent with estimates for coronal heating in active regions derived from observations \citep{Withbroe+Noyes:1977}.

The Ohmic heating rate in other 3D MHD models \citep{Gudiksen+Nordlund:2005a, Gudiksen+Nordlund:2005b, Bingert+Peter:2011} drops (on average) exponentially with height, which is also true when following individual fieldlines \citep{Wettum+al:2013}. These previous models were describing a mature active region with a relatively stable magnetic configuration in which the footpoints are shuffled around. In contrast, in the present model the emerging magnetic field rises into the corona. Thus the interaction between the rising magnetic fieldlines hosting the loop and the ambient magnetic field also contributes to the currents along the loop, so that the Ohmic heating rate is quite constant along the loop (\fig{F:ener}b). The viscous heating is almost two orders of magnitude smaller, so that the Ohmic heating dominates the energy input. 

The heat conduction term is negative near the apex, i.e. it transports the energy added by the Ohmic heating to the lower part of the loop.  Ultimately, the energy is radiated close to the footpoints where the temperature is low.

The advection term at normalized arc lengths of 0.15 to 0.3 (and symmetric on the other side of loop) demonstrates the evaporation upflow filling the loop (\fig{F:loop}d). This converging flow towards the loop top provides compressional work adding energy near the loop apex. This compressional work nearly equals the Ohmic heating at the loop top.

The effect of all contributions, i.e. the right-hand side of \equ{E:ener}, is positive. This leads to a net rise of $\partial e_{\rm th} / \partial t$ of the order of $e_{\rm{th}}/\tau{\approx}10^{-3}$ W m$^{-3}$ (see \fig{F:ener}b). In the coronal part of the loop the number density is about  $n{\approx}10^9$ cm$^{-3}$. Therefore the required increase of the energy $e_{\rm{th}}{=}\frac{3}{2}nk_{\rm{B}}T$  to reach coronal temperatures of about $T{\approx}10^{6}$ K is of the order of $\tau{\approx}1$\,min. This time is compatible with the synthesized images, in which we see the loop forming in a matter of minutes (see animation attached to \fig{F:em}).

In their 2D study \cite{Hurlburt+al:2002} implicitly assumed that the corona adjusts instantly to changes in the heat input because they employ a series of (static) equilibrium models. Here we see that the time scale for the evolution of the loop (minutes) is comparable to the time scale of the energy injection from the photosphere through the Poynting flux (see also \sect{S:trigger}). Thus one has to account for the evolution of the thermal properties in a dynamic model.

\subsubsection{Cooling phase\label{S:loop_ener_c}}

After the heating ceases the loop enters a slow cooling phase (\fig{F:ener}c). Owing to the drop of  Ohmic heating, the plasma pressure falls, the plasma loses its support, and the loop drains. Consequently, decompression is the dominating cooling agent at the apex, as is illustrated by the negative contribution of the compression work throughout the top half of the loop (\fig{F:ener}c). Along with the draining, advection  transports energy from the loop top to the lower parts. In this late phase the dominant heating of the apex is due to heat conduction from the sides. Potentially, such situations can lead to a loss of equilibrium and catastrophic cooling \citep{Muller+al:2003, Muller+al:2004}, which we do not observe here because the heating is not concentrated strongly enough towards the footpoints.

In 3D models with a more stable magnetic field configuration, the loop can reach a (quasi-) equilibrium state, and remain stable for a longer time \citep{Peter+Bingert:2012}. In our model, the magnetic field is expanding due to the flux emergence. Thus the loop cannot reach a (quasi-) equilibrium state and remains a transient feature evolving fast on a time scale of much less than 30 min (see animation with \fig{F:em}).
This is underlined by the fact that the main cooling agent (in the top part) is decompression of the plasma driven by the expansion of the magnetic field.

\begin{figure}
\includegraphics{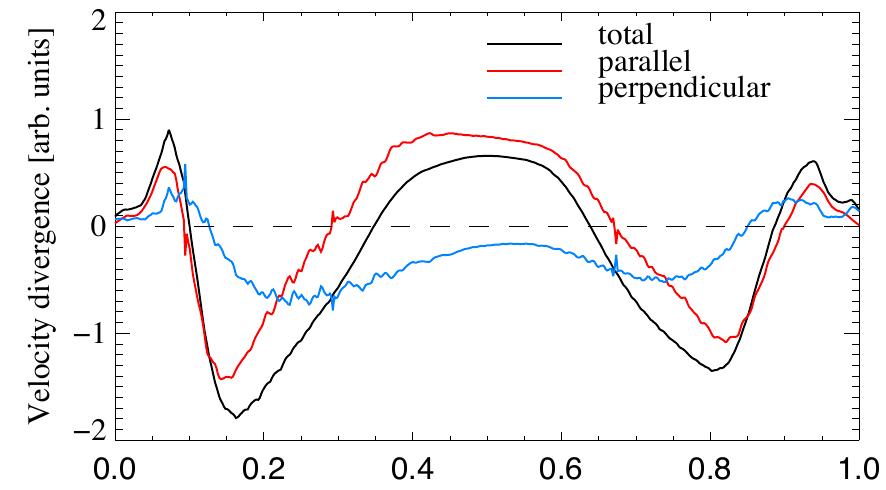}
\caption{Divergence of velocity along the loop at $t{=}14.5$\,min during the loop formation phase. The total divergence $-\nabla \cdot \vec{u}$ is shown in black, the parallel component in red, and the perpendicular component in blue. The arc length along the fieldline is normalized to the loop length. See \sect{S:compression}. 
\label{F:div}}
\end{figure}

\subsubsection{Perpendicular compression}\label{S:compression}
A static rigid 1D loop model only accounts for the compression work from velocity parallel to the magnetic fieldline. However, the compression or expansion perpendicular to the fieldline contributes to the thermal energy density in a 3D model, in particular, if the loop is expanding and interacting with the ambient magnetic field, as it is the case here. 

To evaluate the role of the perpendicular compression, we split the divergence of the velocity, $\nabla \cdot \vec{u}$, into its parallel component ($\left(\vec{b} \cdot \nabla \right)u_{\parallel}$) and its perpendicular component. The latter is  evaluated by $\nabla \cdot \vec{u}-\left(\vec{b} \cdot \nabla \right)u_{\parallel}$.  \fig{F:div} shows these contributions at $t{=}14.5$\,min, i.e., during the loop formation phase. For consistency with \fig{F:ener} we plot $-\nabla \cdot \vec{u}$. A positive value in \fig{F:div} implies convergence/compression, and a negative one implies divergence/expansion. Near the loop top, the parallel contribution shows a converging pattern, because evaporation flows from loop footpoints meet at the loop top (cf., \fig{F:loop}). In contrast, the perpendicular contribution shows a diverging pattern at the top. This is corresponding to the expansion of the magnetic tube which will be discussed in \sect{S:sect_evo}. Still, the net effect is a compression of the plasma. Also in the lower part of the loop the total divergence is basically determined by the parallel contribution.

Although the perpendicular divergence has non-negligible contribution throughout the loop,  the profile of the total divergence mostly follows that of the parallel contribution. This suggests that the 1D description of the flow in the loop is still acceptable at this stage.  However, later the loop shows a clear 3D nature, which is discussed in the next section.

\section{The 3D nature of the loop\label{S:sect}}

\begin{figure}
\includegraphics[width = 9 cm]{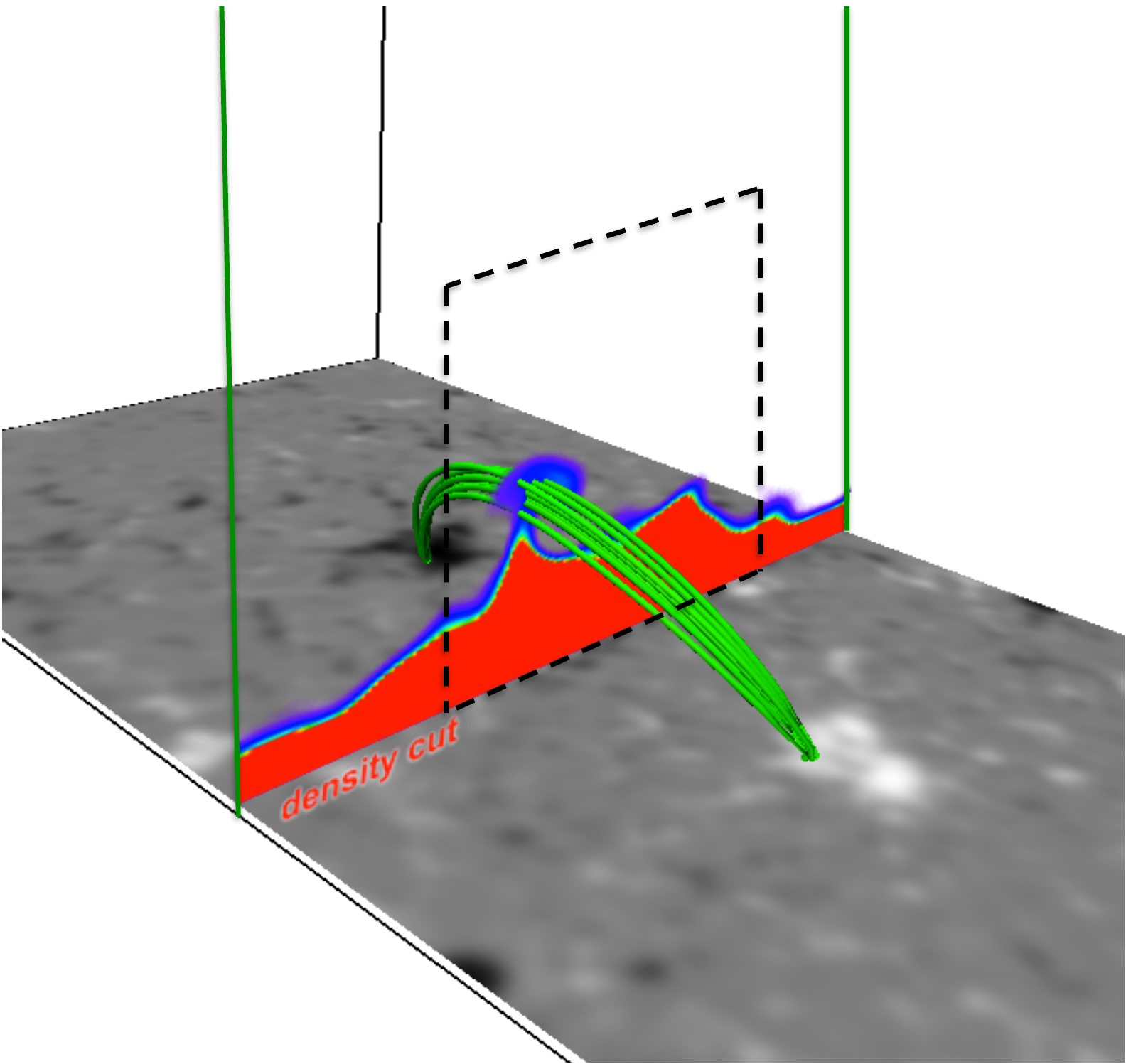}
\caption{3D Visualization of the location of the bright loop at time $t{=}14$\, min. The green lines show the fieldlines roughly outlining the volume of the first bright loop appearing in the simulation. The bottom plane shows the vertical component of the photospheric magnetic field. A density cut in the vertical plane halfway between the footpoints of the loops perpendicular to the loop is indicated by the large green square. On this cut red indicates chromospheric densities, and blue enhanced coronal densities of about $10^9$\,cm$^{-3}$. Lower densities in the corona are transparent.
The black square in dashed line on the midplane indicates the field of view in \fig{F:sect}. 
See \sect{S:sect}. 
\label{F:sect_cut}}
\end{figure}

\subsection{Evolution of the magnetic envelope\label{S:sect_evo}}

To study the magnetic envelope of the EUV loop seen in the synthesized \ang{193} images, we investigate the evolution of a magnetic tube that is (at one particular time) roughly co-spatial with the volume of the EUV loop. We define the magnetic tube based on a \emph{vertical} cut perpendicular to the loop plane in the middle between the two sunspots ($x{=}73$\,Mm) at the time $t{=}14$\,min (see \fig{F:sect_cut}). On this plane, we choose several points roughly enclosing the cross section of the synthesized AIA \ang{193} loop as starting points to follow magnetic fieldlines. This set of fieldlines defines the magnetic tube that we study further. We follow the magnetic tube in time by the same method as used in \sect{S:loop} and investigate the evolution of the cross section of the tube in the vertical midplane between the loop footpoints (large green square in \fig{F:sect_cut}).

\begin{figure}
\includegraphics{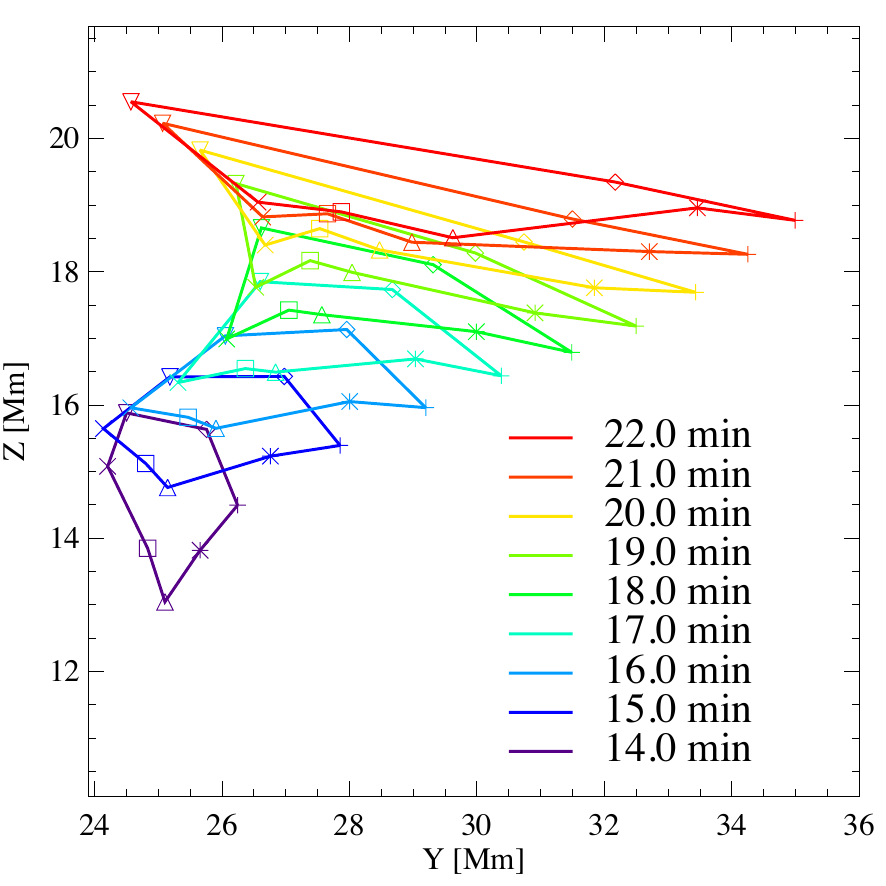}
\caption{Evolution of cross section of the magnetic tube roughly encompassing the  loop appearing in the synthesized AIA \ang{193} images (cf.\ \fig{F:em}). The symbols show the positions of the fieldlines used to define the magnetic tube in the vertical midplane between the loop footpoints (cf.\ green square in \fig{F:sect_cut}). The same symbols indicate the same fieldline at the times color coded according to the legend. See \sect{S:sect_evo}.
\label{F:sect_evol}}
\end{figure}

We depict the temporal evolution of the cross section of the magnetic tube in the vertical midplane in \fig{F:sect_evol}. The magnetic tube moves upward as a whole and the cross section is significantly deformed. From $t{=}14$\,min to $t{=}22$\,min the cross section contracts in the vertical direction and expands significantly in the horizontal direction. This appearance of the magnetic tube is consistent with the rise of the synthesized AIA \ang{193} loop spine in the vertical direction and its significant horizontal expansion (see \fig{F:em} and attached animation). An oblate shape of flux tubes was recently also reported by \citet{Malanushenko+Schrijver:2013} who analyzed the cross section of thin flux tubes in a potential field model. They found that the cross section is distorted for the end-to-apex mapping. That the magnetic tubes in the corona will be non-circular in cross section has already been reported before \citep{Gudiksen+Nordlund:2005b}.

\subsection{Fragmentation of the loop\label{S:sect_frag}}

\begin{figure}
\includegraphics{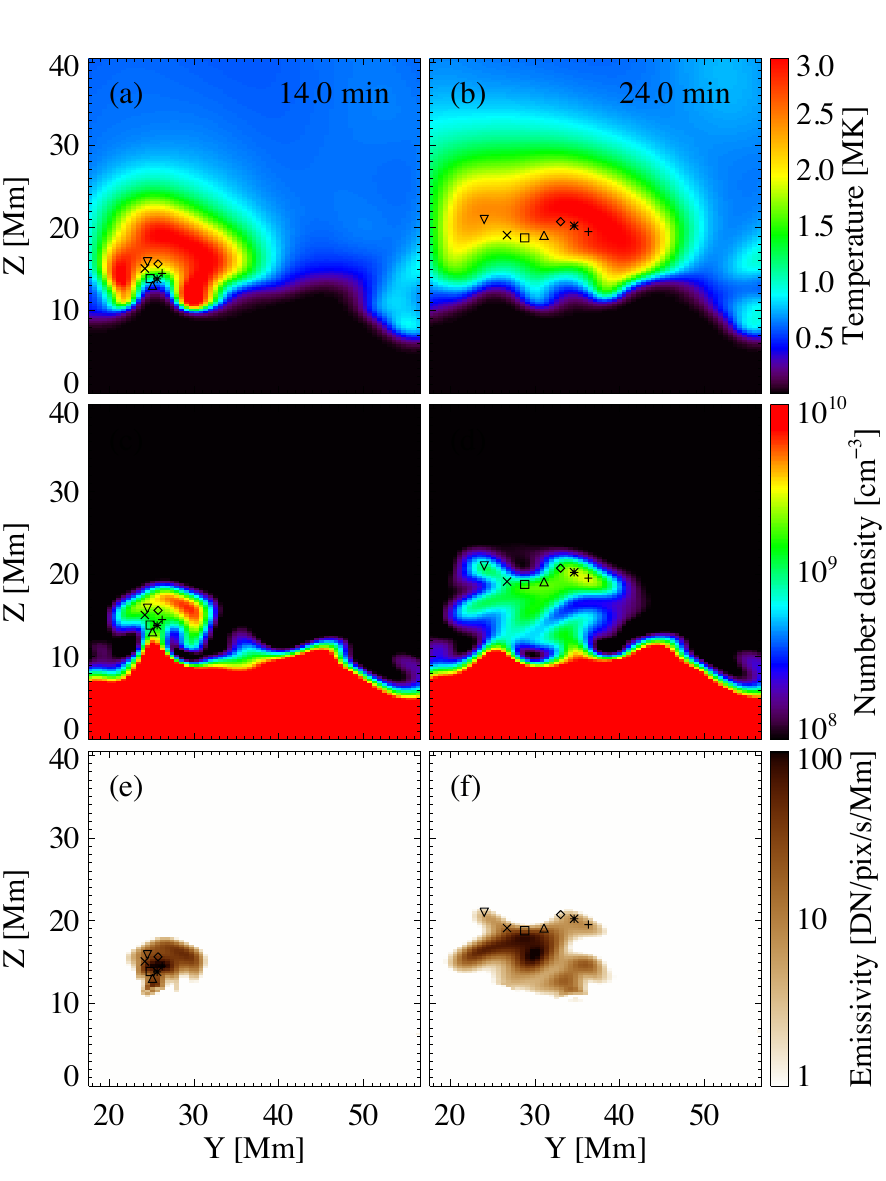}
\caption{Temperature, density and synthesized AIA \ang{193} emission in a vertical midplane between the loop footpoints (cut at $x{=}73$\,Mm). The left and right column show snapshots 10\,min apart at the times indicated in the top panels. In the temperature plots green roughly represents the temperature of maximum contribution to the \ang{193} channel. The black symbols  indicate the cross section of the magnetic tube discussed in \sect{S:sect_evo} and \fig{F:sect_evol}. The field of view roughly matches the black square in dashed line in \fig{F:sect_cut}. See \sect{S:sect_frag}.
\label{F:sect}}
\end{figure}

In \fig{F:em} (and the attached animation) one can see that the synthesized AIA \ang{193} loop is a thin bright structure at the early stage, and then expands. The single bright loop breaks into several individual strands, which is best seen in the top view of the box (\fig{F:em}b). We use the term \emph{fragmentation} for this process. Inspecting the temporal evolution in the movies attached to \fig{F:em} it is clear that this fragmentation means that the original loop fades and new fragments or strand continuously form and dissolve, giving the overall impression of a fragmentation. So this fragmentation is not to be understood in a way as a piece of wood would splinter, but as a coming and going of strands in a growing envelope.

To investigate this process, we show in \fig{F:sect} vertical cuts through the box in the midplane between the two footpoints. This midplane is roughly perpendicular to the loop (same plane as discussed above, cf.\ \fig{F:sect_cut}).
The fragmentation of the loop is visible in the coronal emission emerging from this plane (bottom row of \fig{F:sect}). At the later stage (\fig{F:sect}f) individual patches of AIA \ang{193} emission have formed  that would correspond to individual strands of the larger envelope.

To understand this EUV loop fragmentation we have to investigate the temperature and density structure in the vertical plane. For this we show in \fig{F:sect} also the temperature (top) and density (middle) in the vertical midplane. During the 10 min between the two snapshots shown in the left and right column, the temperature and density structures move upward, and expand horizontally. The density structure looks less smooth than the temperature which is in part because of the draining and filling of the corona.

The \ang{193} emissivity (bottom row) is the product of the density squared and the temperature response function for that channel \citep{Boerner+al:2012}. The latter largely (but not only) reflects the  contribution function of \ion{Fe}{12}, which is strongly peaked with a maximum near 1.5\,MK. In effect, the strong \ang{193} emission originates from locations where the density is high and the temperature is near the peak of the response function for this particular channel. Consequently, the \ang{193} emission pattern is neither cospatial with the density nor with the temperature structure, as is also clear from comparing the panels in the right column of \fig{F:sect} at the later time. The emission structure appears to be much more fragmented than both the temperature and density structure. This is simply because the density and temperature structures are not cospatial, and thus the convolution of the (smooth) density and temperature structures leads to the more clumpy  coronal \ang{193} emission. The same is also true for the other AIA coronal channels, which we do not show here.  

As a side remark, we find in this work a temperature gradient perpendicular to the loop spine with an increasing temperature with height (from about $z{=}14$\,Mm to 22\,Mm). This is similar to the model of \citet{Peter+Bingert:2012} who proposed a new mechanism to explain the constant cross section of coronal loops. Thus some parts of the high density structure at higher temperatures are cut off by the temperature response (or contribution) function, and in EUV emission the loops looks as if having a constant cross section, even though the plasma loop, i.e. the density structure, expands along the loop, or more precisely, with the magnetic tube.  Even though we do not investigate this further in detail here, the \ang{193} loop shown in \fig{F:em} from the side has roughly constant cross section. This is based on the same process as outlined by \citet{Peter+Bingert:2012}.

\section{What triggers the loop formation?\label{S:trigger}}
The appearance of the model corona is compatible with EUV observations in the sense that a clearly distinguishable loop forms in the synthesized images. The question remains why the loop forms at that particular time and position. We investigate this by checking the energy input into the loop which is given through the Poynting flux, $\tilde{\vec{S}} = \left( \eta\vec{j} - \vec{u} \times \vec{B} / \mu_0 \right) \times \vec{B}$. Near the bottom boundary, the driving by the photospheric convective motions from the flux-emergence model induces strong currents, which are mainly confined to the bottom layers. The amplitude of the resistive term, $\eta\vec{j}$ drops very fast with height and becomes much smaller than the $\vec{u} \times \vec{B} / \mu_0$ term, in particular in the area near the simulated spots. Thus when studying the energy input into the coronal part of the loop it is sufficient to investigate the  $\vec{u}
\times \vec{B}$ part alone,
\begin{equation}\label{E:poynting}
\vec{S} = -\frac{1}{\mu_0} \left( \vec{u} \times \vec{B} \right) \times \vec{B}~.
\end{equation}

\begin{figure}
\includegraphics{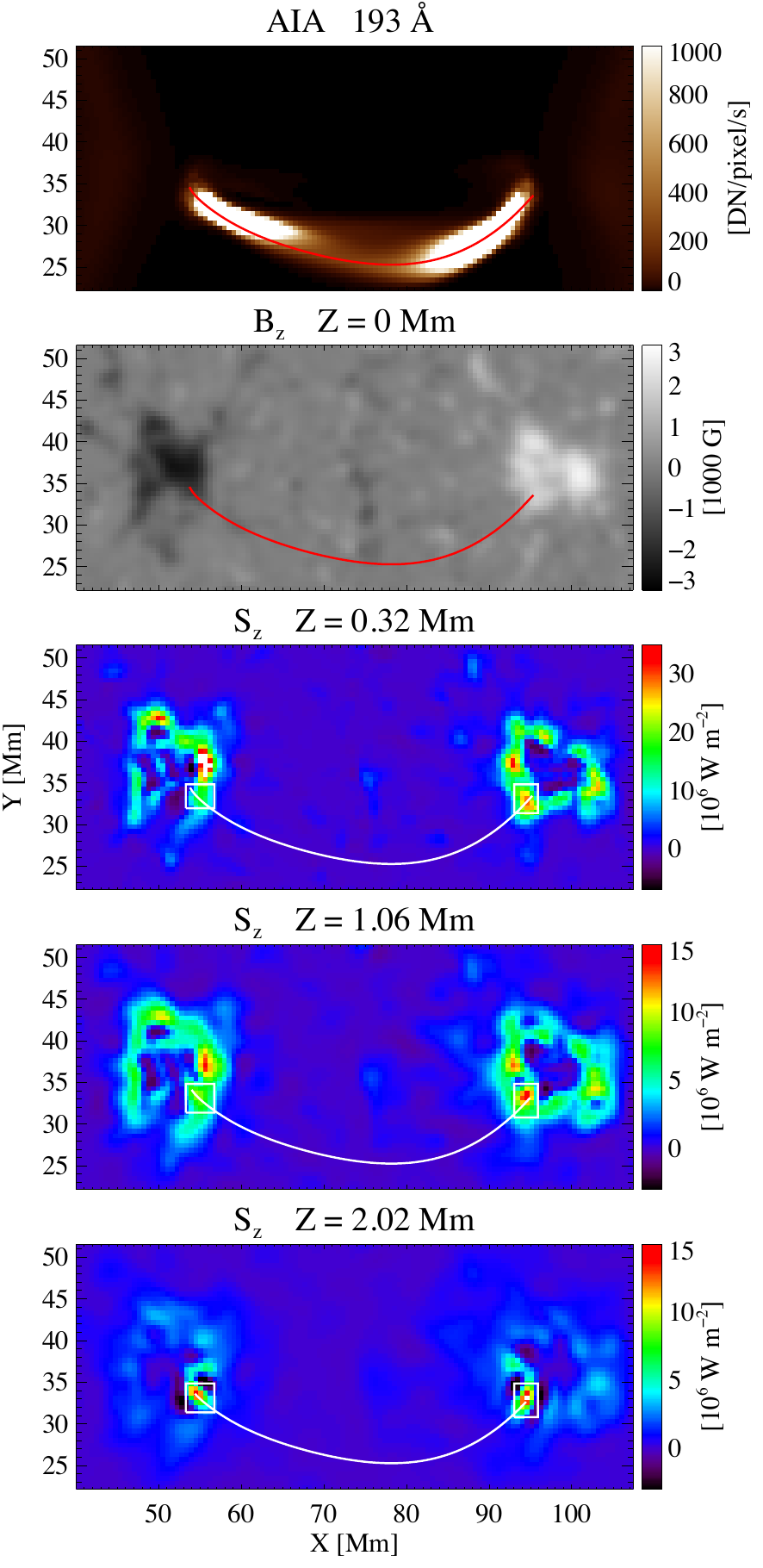}
\caption{Zoom into the emerging active region at $t{=}14$\,min. The top two panels show the synthesized AIA \ang{193} emission integrated along the vertical and the magnetic field, $B_z$, at the bottom boundary. The lower three panels show the vertical component of the Poynting flux, $S_z$, at three heights. The red and white lines indicate the same magnetic fieldline at the spine of the loop as shown in \figs{F:em}. The white boxes around both footpoints in the lower panels indicate the regions where we calculate the average vertical Poynting flux in \fig{F:sz_curve}.
\label{F:poyn}}
\end{figure}

In \fig{F:poyn} we show the vertical component of the Poynting flux $\vec{S}$ in horizontal slices at three heights, from the photosphere ($z{=}0.32$\,Mm) to the coronal base ($z{=}2.02$\,Mm). There is a clear enhancement of the upward-directed Poynting flux surrounding the sunspot areas forming sort of a ring around the sunspot (green in \fig{F:poyn} for $z{=}0.32$\,Mm). This enhancement is at least a factor of five to ten with respect to the surrounding quiet Sun area or the center of the sunspot. In the former the magnetic field is too weak, in the latter the strong magnetic field suppresses the horizontal motions, so that in these regions no considerable upward directed Poynting flux can be found. This is consistent with the widely known observational fact that coronal loops in EUV and X-rays do not originate from the center of sunspots where the magnetic field is the strongest, but from the periphery of sunspots, i.e. the outer parts of the penumbra. In our model this is reflected by the fact that only in the periphery the upward Poynting flux is significantly enough to power coronal loops.

At the \emph{coronal base} ($z{=}2.02$\,Mm)  the Ponynting flux  has the strongest enhancement near both footpoints of the loop, being typically another factor of  about three higher  than in the already enhanced region in the sunspot periphery. In \fig{F:poyn}, this shows up as the red spots in the panel for $z{=}2.02$\,Mm. However, in the photosphere ($z{=}0.32$\,Mm) only the right footpoint shows an enhancement of the Poynting flux, but not the left one. A closer inspection  at the bottom boundary shows that this enhancement near the right footpoint in the \emph{photosphere} is due to small magnetic flux elements which are advected by the convective motions into the strong magnetic field of the sunspot. These magnetic flux elements have sizes of ${\approx}$3\,Mm, which is the scale of energy input into the loop and is not too far from the smallest resolvable scale in this model.

We miss a lot of the small-scale motions and fine magnetic structures in the photosphere when we map the original flux-emergence simulation to the grid of the coronal simulation (see \sect{S:model_coupling}). This can have two consequences. Firstly, the energy input into the corona in our model is reduced, because we miss the Poynting flux on these smaller scales, at most this is a factor of two. Because the temperature scales with the energy input to the power of 2/7 \citep{Rosner+al:1978}, this would have only a minor impact on the temperature, but it might be that the coronal density in our model is too low by up to a factor of 2 in some places. Secondly, the higher spatial resolution in the photosphere, properly resolving granulation, will give rise to finer structures in the corona, too. These conclusions are supported by the preliminary results from a high-resolution numerical experiment.

\begin{figure}
\includegraphics{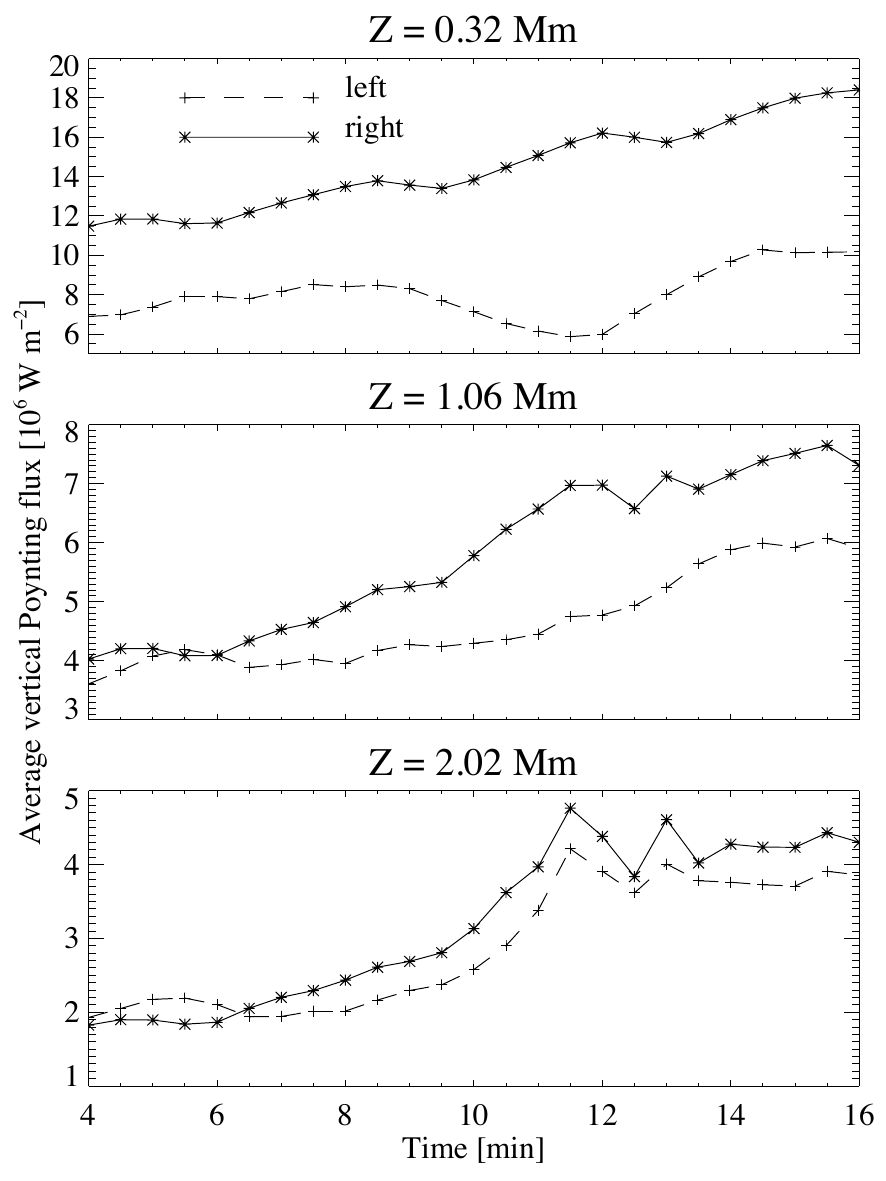}
\caption{Temporal evolution of the vertical Poynting flux at the loop footpoints at three different height from the surface (top panel) to the coronal base (bottom panel).  Here we show the averages in horizontal regions around the fieldline indicated in \fig{F:poyn} by the respective boxes. The dashed lines are for the left footpoint, and solid lines are for the right footpoint. 
\label{F:sz_curve}}
\end{figure}

To further investigate the vertical Poynting flux at the footpoints of the loop we study the temporal variation of the vertical Poynting flux in different heights along the loop. We do this in terms of averages in a small horizontal section around the loop as indicated by the rectangles in \fig{F:poyn}. The sizes of the rectangles are slightly different for the left and right footpoints and for different heights in order to best capture the Poynting flux enhancement. The positions of the rectangles are fixed in time. The resulting averages as a function of time are plotted in \fig{F:sz_curve}.

At $z{=}0.32$\,Mm, the vertical Poynting flux at the right footpoint (solid) increases significantly by more than 6\,MW/m$^2$ during the 12\,min shown in the plot around the time the loop appears (\fig{F:sz_curve}). A Fourier analysis clearly shows that this increase is modulated with a timescale of about 4\,min, which is close to the 5-minutes oscillations in the photosphere and close to the lifetime of granules. In contrast, the left footpoint shows no significant increase over time, but only the granulation modulation. That the left and the right footpoint show a different behaviour in the \emph{photosphere} is not surprising, because in the flux emergence simulation these two footpoint regions, which are located in the different sunspots, evolve independently.

The situation is different higher up in the atmosphere. At $z{=}1$\,Mm
and 2\,Mm both footpoints show a significant increase, with  the the right footpoint preceding the rise of the left footpoint at both heights.  

Based on the timing shown in \fig{F:sz_curve} one can sketch the following scenario. At the \emph{right} footpoint in the low photosphere the upward Poynting flux is increasing because of the near-surface convection driven by the flux emergence simulation. This disturbance of the field then travels upward through the high-plasma-$\beta$ region and can be seen in the upper photosphere near $z{=}1$\,Mm and further propagates upward to the coronal base at $z{=}2$\,Mm, where plasma-$\beta$ is below unity. Here we can see a steeper rise after the magnetic stresses have been built up slowly from below. We also see a clear rise on the \emph{left} footpoint at the coronal
base at $z{=}2$\,Mm. However, this increase lags behind the rise in the \emph{right} footpoint by some 30\,s, which is close to the Alfv\'en crossing time (with a loop length above   $z{=}2$\,Mm of about 40\,Mm and an average Alfv\'en speed of about 2000\,km/s in the coronal part). This underlines that the magnetic disturbance travels from the right coronal base to the left coronal base and triggers there a perturbation, that in the end leads to an increased Poynting flux also on the left side. From the left footpoint at $z{=}1$\,Mm
we can see that this disturbance can penetrate a bit into the high-$\beta$ region, but cannot reach all the way down into the photosphere (to $z{=}1$\,Mm). This is also because of the strong density stratification.

In conclusion, 
the time profiles of the Poynting flux at different heights imply that the enhancement at one (right) footpoint near the bottom induces the increase of Poynting flux in higher layers on the same side. This also  induces an increase of the Poynting flux on other side of the loop at the coronal base, but not down to the photosphere. As a consequence of the similarly increased Poynting fluxes on \emph{both} sides at the coronal base, the heat input into the loop is comparably symmetric as already discussed in \sect{S:loop_ener}.

Other loops that  form later show similar features. A further numerical experiment with increased spatial resolution will have to show if this result can be substantiated. In particular, this will have to investigate to what extent the small-scale evolution of the (inter-)granular magnetic fields can make their way up into the corona and thus alter the spatio-temporal evolution of the Poynting flux in both loop footpoints at the coronal base.

 \section{Summary}
In this paper, we presented a coronal model of an emerging active region driven by a simulation of magnetic flux emergence from the convection zone through the photosphere. The magnetic field  expands  into the corona, while a pair of simulated spots forms in the photosphere. Ohmic dissipation heats the coronal plasma, while heat conduction along magnetic fieldlines, radiative losses through optically thin radiation, and flows  carry away the energy input. The  treatment of the full energy balance ensures that the coronal pressure is set self-consistently and allows us to synthesize the  EUV emission from the model corona. 

Once sufficient magnetic flux was emerged through the surface and the coalescence of small-scale magnetic patches formed large-scale magnetic patches turning into sunspots, the first EUV coronal loops form within minutes. The EUV  loop rises upwards, expands significantly in the horizontal direction, and, most importantly, fragments into several individual EUV structures, i.e. the changing heat input produces new strands in a growing envelope.

The energy input is driven by the advection of the magnetic field in the photosphere, i.e. by the horizontal convective motions. Connected by magnetic fieldlines through the corona, the regions of enhanced Poynting flux at one end induce an increase of the Poynting flux at the other end at the coronal base. The upward directed Poynting flux leads to an increased energy input giving rise to the heating of the coronal plasma and the enhancement of the pressure due to the evaporative upflows.  The emerging magnetic field hosting the forming loop rises into the ambient magnetic field and currents also build up near the upper part. These contribute to the Ohmic heating   in the top part of the loop leading to a nearly constant heat input along the loop. 

In its early evolution  the coronal loop behaves (at least concerning the energy budget) similarly as a conventional 1D loop model would predict if we would prescribe the energy input. However, in the later stages the loop shows its true 3D nature. The horizontal magnetic expansion and in particular the fragmentation of the EUV emission are a clear indication that a 1D model would not be sufficient to describe a newly forming emerging loop. In  the cross-sectional cut perpendicular to the EUV loop, the temperature and the density structure are comparably smooth but not exactly cospatial. This gives rise to the fragmented appearance of the loop in EUV emission with threads (or loop-fragments, or strands) with diameters much smaller than the typical spatial structures in temperature or density.

Our model of the formation and evolution of a EUV coronal loop in an emerging active region sheds new light on our understanding of coronal loop formation. A further analysis of this and more advanced numerical experiments will have to investigate the differences (and similarities) of the evolution of coronal loops seen in different wavelength bands, in particular towards X-rays, and how the forming loops would appear in spectroscopic observations. Of particular interest will be the further investigation of the evolution of the magnetic field structure in relation to the spatial structure of the synthesized coronal emission. 

\begin{acknowledgements}
{We thank M.~Sch\"ussler and R.~Cameron for discussions and comments that improve this paper. We gratefully acknowledge the constructive comments by the anonymous referee. This work was supported by the International Max-Planck Research School (IMPRS) on Physical Processes in the Solar System and Beyond. This work was partially funded by the Max-Planck/Princeton Center for Plasma Physics. The computations were done at GWDG in G\"ottingen and SuperMUC. We acknowledge PRACE for awarding us the access to SuperMUC based in Germany at the Leibniz Supercomputing Centre (LRZ). The visualization in \fig{F:3d} and \fig{F:sect_cut} were done using VAPOR ({www.vapor.ucar.edu}). }
\end{acknowledgements}

\bibliographystyle{aa}
\bibliography{reference}

\begin{appendix}
\section{Energy budget along a loop in one dimension}\label{appen}
To investigate the energy budget along the loop structure we made the following assumptions to derive \equ{E:ener}.
Under such assumptions, the right-hand-side in \equ{E:ener} basically describes the evolution of the energy along a (field) line that moves with the velocity perpendicular to the line itself.

\begin{enumerate}
\item \emph{Constant cross section}.\\
{$\vec{B}$ is invariant along the magnetic tube, which is the same as to say that the loop has a constant cross section. Therefore, the unit vector of the magnetic field, $\vec{b}$, satisfies 
$$
\nabla \cdot \vec{b}~=~\frac{1}{\abs{\vec{B}}}\nabla \cdot \vec{B}+\vec{B} \cdot \nabla \frac{1}{\abs{\vec{B}}}~=~0 ~.
$$
}

\item \emph{No compression perpendicular to the loop}.\\
{The velocity is decomposed into $u_{\parallel} \vec{b}~+~u_{\perp}\boldsymbol{\delta}$, with the parallel $u_{\parallel}=\vec{u}\cdot\vec{b}$, the perpendicular component $u_{\perp}=\abs{\vec{u}-u_{\parallel}\vec{b}}$, and $\vec{b}\cdot\boldsymbol{\delta}=0$. We assume the magnetic tube is not compressed by the flows in the perpendicular direction, which implies $\nabla \cdot \left(u_{\perp}\boldsymbol{\delta} \right)=0.$ This assumption is appropriate for a rigid 1D loop model, although it is not the case in our 3D simulation. }

\item \emph{Heat conduction parallel to the loop}.\\
{The heat flux is along the magnetic field, i.e. $\vec{q}=q_{\parallel}\vec{b}$. Because $B$ is invariant along the magnetic tube, 
$$
\nabla \cdot \vec{q} = \vec{b}\cdot\nabla q_{\parallel}~+~q_{\parallel}\nabla\cdot\vec{b}=\vec{b}\cdot\nabla q_{\parallel}.
$$
}
\end{enumerate}

These assumptions are consistent with traditional 1D loop modelling. They do not fully hold in the 3D loop we find in our numerical experiment, but are appropriate for the purpose of the comparison made in \sect{S:loop}.

In general, the conservation of thermal energy is written as
\begin{align}
\frac{\partial e_{\rm th}}{\partial t}  =& - \left(\vec{u}  \cdot \nabla  \right) e_{\rm th} - \frac{\gamma}{\gamma-1}p\left(\nabla\cdot \vec{u}\right)~+~Q~-~L~-~\nabla \cdot \vec{q} ~.
\end{align}

With the above assumptions this energy budget can be rewritten as\\
\begin{align}
\nonumber
\frac{\partial e_{\rm th}}{\partial t}  =& - u_{\parallel}\left( \vec{b}  \cdot \nabla  \right) e_{\rm th} - u_{\perp}\left(\boldsymbol{\delta}  \cdot \nabla  \right) e_{\rm th} \\
\nonumber
& - \frac{\gamma}{\gamma-1}p\bigg[\nabla\cdot \left(u_{\parallel} \vec{b} \right)+\nabla\cdot \left(u_{\perp}\boldsymbol{\delta}\right)\bigg] \\
&~+~Q~-~L~-~\vec{b}\cdot\nabla q_{\parallel},
\end{align}
\begin{align}
\nonumber
\frac{\partial e_{\rm th}}{\partial t}  =& - u_{\parallel}\left( \vec{b}  \cdot \nabla  \right) e_{\rm th} - u_{\perp}\left(\boldsymbol{\delta}  \cdot \nabla  \right) e_{\rm th} - \frac{\gamma}{\gamma-1}p\left(\vec{b} \cdot \nabla \right) u_{\parallel}\\
&~+~Q~-~L~-~\vec{b}\cdot\nabla q_{\parallel},
\end{align}
where the definitions of $Q$, $L$, $q_{\parallel}$ are as in \equ{E:ener}. We move the term related to $u_{\perp}\boldsymbol{\delta}$ to the left-hand-side of the equation and define
\begin{equation}
\left( \frac{\partial e_{\rm th}}{\partial t}\right)_s
~~=~~ \left[ \frac{\partial e_{\rm th}}{\partial t}+ u_{\perp}\left(\boldsymbol{\delta}  \cdot \nabla  \right) e_{\rm th} \right] ~. 
\end{equation}
This can be considered as a type of material derivative.

With this the energy budget reads, 

\begin{align}
\nonumber
\left(\frac{\partial e_{\rm th}}{\partial t}\right)_s  =& - u_{\parallel}\left( \vec{b}  \cdot \nabla  \right) e_{\rm th} - \frac{\gamma}{\gamma-1}p\left(\vec{b} \cdot \nabla \right) u_{\parallel}\\
&~+~Q~-~L~-~\vec{b}\cdot\nabla q_{\parallel} ~,
\end{align}
which is just \equ{E:ener}.

\end{appendix}

\end{document}